\documentclass[a4paper,authoryear,10pt]{elsarticle}
\usepackage{color,graphicx,natbib,amssymb}
  
\usepackage{hyperref}
\hypersetup{
 colorlinks  = true, 
 urlcolor    = blue, 
 linkcolor   = blue, 
 citecolor   = blue  
 }
\usepackage{tikz,xcolor,hyperref}

\definecolor{lime}{HTML}{A6CE39}
\DeclareRobustCommand{\orcidicon}{
	\begin{tikzpicture}
	\draw[lime, fill=lime] (0,0) 
	circle [radius=0.16] 
	node[white] {{\fontfamily{qag}\selectfont \tiny ID}};
	\draw[white, fill=white] (-0.0625,0.095) 
	circle [radius=0.007];
	\end{tikzpicture}
	\hspace{-2mm}
}
\foreach \x in {A, B, C, D}{\expandafter\xdef\csname orcid\x\endcsname{\noexpand\href{https://orcid.org/\csname orcidauthor\x\endcsname}
			{\noexpand\orcidicon}}
}

\begin{document}

\newcommand{\arcm}{$^\prime$}
\newcommand{\arcs}{$^{\prime\prime}$}
\newcommand{\m}{$^{\rm m}\!\!.$}
\newcommand{\D}{$^{\rm d}\!\!.$}
\newcommand{\F}{$^{\rm P}\!\!.$}
\newcommand{\kms}{km~s$^{-1}$}
\newcommand{\ks}{km~s$^{-1}$}
\newcommand{\ms}{M$_{\odot}$}
\newcommand{\rs}{R$_{\odot}$}
\newcommand{\oc}{$O\!-\!C$}
\newcommand{\ubv}{\hbox{$U\!B{}V$}}
\newcommand{\bv}{\hbox{$B\!-\!V$}}
\newcommand{\ub}{\hbox{$U\!-\!B$}}

\newcommand{\ond}{Ond\v{r}ejov}
\newcommand{\ova}{Ostrava}
\newcommand{\hip}{$Hipparcos$}
\newcommand{\valmez}{Vala\v{s}sk\'e Mezi\v{r}\'{\i}\v{c}\'{\i}}
\newcommand{\stef}{\v{S}tef\'anik Observatory}
\newcommand{\vv}{V608~Cam}

\def\astrobj#1{#1}

\begin{frontmatter}

\title{A photometric study of \vv: \\ 
apparent period changes as a result of surface activity}

\author[label1,label2]{F. \v{S}ebek}
\ead{filip.sebek.fs@gmail.com}
\author[label2,label3]{F. Walter\orcidB{}}
\author[label4]{M. Wolf}

\address[label1]{High School Boti\v{c}sk\'a, Boti\v{c}sk\'a 1, CZ-128~01 Praha 2, Czech Republic}
\address[label2]{Czech Astronomical Society, Variable Star and Exoplanet Section,           V\'ide\v{n}sk\'a~1056, CZ-142~00~Praha~4, Czech Republic}
\address[label3]{Stefanik Observatory Prague, Strahovsk\'a~205, 
          CZ-118~00~Praha~1, Czech Republic}
\address[label4]{Astronomical Institute, Faculty of Mathematics and Physics, Charles University, V~Hole\v{s}ovi\v{c}k\'ach~2, CZ-180~00~Praha~8,
   Czech Republic}

\begin{abstract} 
The VRI light curves were measured for the low-mass eclipsing binary \vv\ as a part of our long-term observational project for studying of eclipsing binaries with a short orbital period.
The {\sc Tess} light curve solution in {\sc Phoebe} results to the detached configuration, where 
the temperature of primary component was fixed to $T_1 = 5\,300 $ K according to {\sc Gaia} results, which gives us $T_2 = 4\,110 \pm 50 $ K for the secondary. The spectral type of the primary component was derived to be K0 and the photometric mass ratio was estimated $q = 0.92 \pm 0.07$. Characteristics and temporal variation of the cold region on the surface of the secondary component were estimated and are attributed to apparent period changes of this eclipsing binary with a cycle of about 2.4~yr. 

\end{abstract}


\end{frontmatter}

\section{Introduction}

Low-mass and late-type stars with masses below 1 \ms\ (spectral types K and M) are the most common and most frequent stars in our Galaxy. 
Long-term photometric monitoring of low-mass eclipsing binaries is a very useful tool for studying star-spot parameters (their structure, coordinates, sizes and temperatures), their evolution and statistical properties. 
Nevertheless, current observations of low-mass stars show a long-lasting discrepancy between estimated and modeled parameters, where the models give 5–10 \% smaller radii than observations (Chabrier \& Baraffe 2000; Morales et al. 2010; Mann et al. 2015). The low-mass stars are also affected by chromospheric activity caused by a strong magnetic field, dark or bright spots. This variable activity has been frequently observed as flares, and plays important role for precise determination of fundamental physical parameters, esp. their radii and temperatures. Their rotation periods are synchronized with the orbital period due to the tidal forces.

Moreover, many low-mass eclipsing binaries display periodic eclipse time variations caused by a third circumbinary component orbiting the eclipsing pair (so-called Light-Time Effect, {\sc Lite}, Irwin 1952) or magnetically-induced modulations caused by an active star in the system (Applegate 1992).
This makes them very promising targets to search for circumbinary brown dwarfs or giant planets by analyzing the LITE. To date, several substellar companions to LMBs have been discovered using this simple but efficient method (Lee et al. 2009; Hinse et al. 2012; 
Pribulla et al. 2012).  

Additional timing variations with small amplitudes might be produced by the asymmetries of the eclipse light curves through stellar activity, such as star spots or small flares. 
The effect of star spots on the \oc\ diagrams was studied by Kalimeris et al.
(2002). They found that star spots modulate the \oc\ values and can introduce high-frequency and low-amplitude disturbances of less than 0.01~d. These variations are caused by the change in the surface density and center of light over the visible hemisphere.
Barros et al. (2013) showed that observed transit time variations in the hot-Jupiter 
WASP-10b system are also due to spot occultation features. Moreover, Korda et al. (2017) tested the spot variability on the light curve of low-mass binaries and found a
difference in mid-eclipse times of about 95~s.
Recently, Zaire et al. (2022) announced, that the well-known eclipsing binary system V471~Tau
has the magnetically active K2 dwarf component, which might be responsible for driving the eclipse timing variations with predicted $\sim 35$~yr activity cycle.

The low-mass eclipsing binary \vv\ (also NSVS~109935, BD+82~160, 
ASASSN-V~J062601.90+822126.3, 1RXS~J062558.2+822124, TYC~4537-0765, 
Sp. K0, $V_{\rm max} = 10.6$ mag) 
is a rather bright but neglected northern object with a short orbital period of about 10~h.  
In the {\sc Simbad}\footnote{\url{http://simbad.u-strasbg.fr/simbad/}}
database \vv\ is also mentioned as a nearby and high proper-motion star. Its variability was discovered by Hoffmann et al. (2008) using the publicly available Northern Sky Variability Survey (NSVS, Wozniak et al. 2004) and classified as a probable low-mass binary with the period of 0.448~d.   

\vv\ is also listed as an object 1RXS~J062558.2+822124 in the ROSAT All-Sky Bright Source Catalogue (Voges et al. 1999) as a faint X-ray source.
For \vv\ the angular diameter of the system $\rho = 7.84 \cdot 10^{-2}$ mas was also measured by the {\sc Gaia} satellite and this value is included in the {\it Mid-infrared stellar Diameters and Fluxes compilation Catalogue} (MDFC ver. 10, Cruzalebes et al. 2019).	 
Recently, \vv\ was included in a study of nearby young stellar association in Cepheus and classified as a spectroscopic binary of SB1 type (Klutsch et al. 2020).
The following linear ephemeris was proposed in the VSX-index\footnote{\url{https://www.aavso.org/vsx/}} for the current use:

\begin{equation}
{\rm Pri.Min. = HJD\; 24\; 57762.8815 + 0.448072} \cdot E.
\end{equation}

\noindent
The {\sc Gaia} DR2 astrometric and phometric data on \vv\ are summarized in Table~\ref{tg} (Gaia Collaboration et al. 2018). The distance to the system was derived to be $d = 115.6 \pm 0.2$ pc (Bailer-Jones et al. 2021).


In this paper we report on a light curve model and surface activity on one component of this low-mass and late-type eclipsing binary. 
To our knowledge, no precise photometric analysis nor period study of \vv\ has been published so far. 
The presented paper is structured as follows. In Section~2, we describe our photometric observations and data analyses.  The improved ephemeris is also derived. 
The new mid-eclipse times and \oc\ diagram is presented in Section~3. 
In Section~4, we analyse \vv\ light curves and  derive photometric parameters of the system.  The discussion of the results is given in Section~5 and our conclusions are presented in Section~6.

\begin{table}
\begin{center}
\caption{{\sc Gaia} DR2 astrometric and photometric data on \vv.}
\label{tg}
\smallskip
\begin{tabular}{cc}
\hline\hline\noalign{\smallskip}
 Parameter              &  Value     \\
\noalign{\smallskip}\hline\noalign{\smallskip}
$\alpha_{2000}$ [h m s] & 06 26 01.76  \\
$\delta_{2000}$ [d m s] & +82 21 27.59 \\
pm $\alpha$ [mas/yr]    & 0.510    $\pm$ 0.020 \\
pm $\delta$ [mas/yr]    & --50.320 $\pm$ 0.016 \\
parallax [mas]          & 8.631    $\pm$ 0.013 \\
\noalign{\smallskip}\hline\noalign{\smallskip}
$B$ [mag]                 & 11.53  $\pm$ 0.07  \\
$V$ [mag]                 & 10.658 $\pm$ 0.013 \\
$G$ [mag]                 & 10.465 $\pm$ 0.007 \\
$BP$ [mag]                & 10.836 $\pm$ 0.013 \\
$RP$ [mag]                & 9.833  $\pm$ 0.013 \\
$J$ [mag]                 & 9.277  $\pm$ 0.020 \\
$H$ [mag]                 & 8.874  $\pm$ 0.015 \\
$K$ [mag]                 & 8.771  $\pm$ 0.020 \\
\hline
\end{tabular}
\end{center}
\end{table}

\section{Observations}

\begin{figure}
\begin{center}
\includegraphics[width=0.85\textwidth]{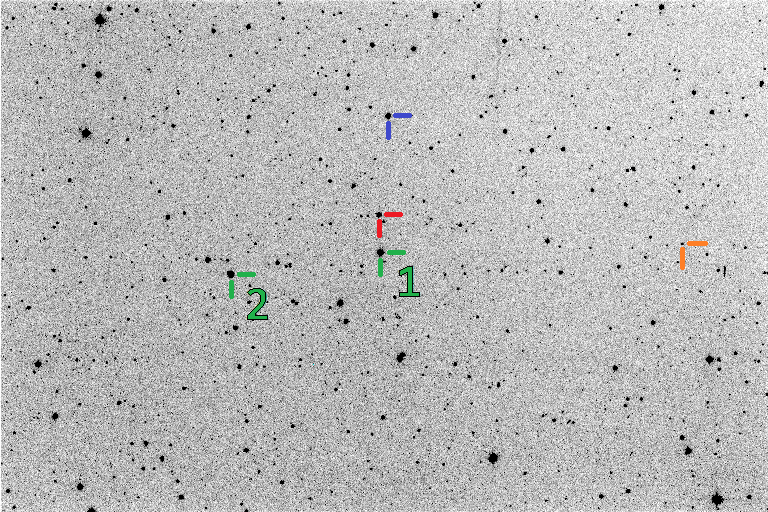}
\caption[ ]{The finding chart of \vv, the dimensions are approx. $1^{\circ}  \times$ 42', north is right. The position of variable (red), comparison 1 and 2 (green) and check (blue) stars are denoted. 
The nearby variable NSVS~108681 is marked orange.} 
\label{fc}
\end{center}
\end{figure}

Since 2020 the time-resolved CCD photometry of \vv, mostly during eclipses, has been regularly obtained at two observatories. 

\begin{figure}[t]
\begin{center}
\includegraphics[width=0.9\textwidth]{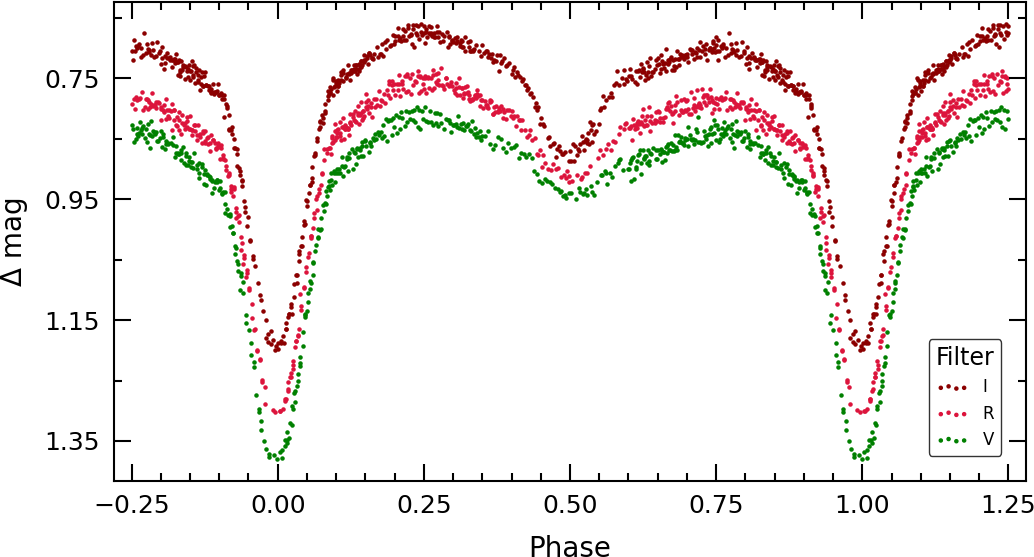}
\caption[ ]{The VRI differential light curves of \vv\ obtained 
            during Sep - Oct 2021 at \valmez.} 
\label{608ValMez}
\end{center}
\end{figure}

\begin{itemize}

\item The initial CCD photometry was performed at \stef\ in Prague, Czech Republic, during Apr -- Sep 2020. The 0.37-m Maksutov-Cassegrain telescope with the CMOS camera ZWO ASI 1600 MM Pro and VRI photometric filters were used. The mean exposure time was 45 sec.

\item The next observations in our campaign were obtained at \valmez\ observatory, Czech Republic, during Sep -- Oct 2021. The 0.15-m Newtonian telescope with remote control, the CCD camera MI G2-1600 and VRI filters were used. 
Individual exposure times lasted up to 60 sec. The folded VRI light curves are shown on Fig.~\ref{608ValMez}.

\item Finally, additional CCD photometry was performed at \valmez\ observatory in March 2022. The 0.15-m Newtonian telescope with remote control, the CCD camera MI G2-1600  and {\sc Sloan} $g, r, i$ filters were used.
Individual exposure times lasted up to 70 sec. 
\end{itemize}

\begin{table}
\begin{center}
\caption{Precise coordinates of selected stars in the field of \vv, see Fig.~\ref{fc}. }
\label{t0}
\smallskip
\begin{tabular}{ccccc}
\hline\hline\noalign{\smallskip}
Object & $\alpha_{2000}$ [h:m:s] & $\delta_{2000}$ [d:m:s] & $V$ [mag] & Other name \\
\noalign{\smallskip}\hline\noalign{\smallskip}      
\vv         & 06 26 01.76  &  +82 21 27.59 & 10.6  &   BD+82 160 \\
comparison1 & 06 27 34.53  &  +82 21 51.04 &  9.55 &   BD+82 161 \\
comparison2 & 06 29 01.16  &  +82 09 53.80 & 10.23 &   BD+82 163 \\
check       & 06 21 57.23  &  +82 21 17.60 & 10.80 &   GSC 04537-00881 \\
\hline
\end{tabular}
\end{center}
\end{table} 

\noindent
Smaller telescopes and modern CCD technique were sufficient for good S/N photometry of such a bright object. Our CCD observations were reduced in a standard way. The images were bias-corrected and flat-fielded before aperture photometry was carried out.
The {\sc C-Munipack}\footnote{Package of software utilities 
for reducing astronomy CCD images, current version 2.1.32, 
available at \url{http://c-munipack.sourceforge.net/}}, a synthetic aperture photometry software, was routinely used.
Time-series were constructed by computing the magnitude difference between the variable and a nearby comparison and check stars, see Fig.~\ref{fc} and Table~\ref{t0}. The heliocentric correction was applied. 
The computer at the \stef\ telescope is synchronized using the local Meinberg NTP
\footnote{\url{https://www.meinbergglobal.com/english/sw/ntp.htm\#ntp_stable}} service using five different NTP time-servers as time reference. 
Resulting local time is smoothed average of differences to all reference servers.
Computer at \valmez\ Observatory is periodically synchronized using a time-server provided by Microsoft. The uncertainties of photometric measurements at two smaller telescopes were always about 0.01 -- 0.02~mag. 

As a northern object with high declination (Dec. $\simeq +82^\circ $), \vv\ was also measured frequently by the  {\it Transiting Exoplanet Survey
Satellite} ({\sc TESS}) in 2-min cadence during several periods (Sectors~19 and 20 in 2019, Sectors~40 and 47 in 2021, and Sector~52 in Dec~2022). These high-quality light curves were used for precise mid-eclipse time determination as well as for modeling of the system.
New times of primary and secondary minima and their uncertainties were generally determined by fitting the light curve by Gaussians or polynomials of the third or fourth order. The least-squares method was used. They are listed in Table~\ref{t1}, where epochs were computed according to the following improved linear ephemeris:

\begin{equation}
{\rm Pri. Min. = HJD\; 24\; 51474.6316(13) + 0.44807272(8)} \cdot E. \\
\label{ephem}
\end{equation}

\noindent
Minima obtained at \stef\ and \valmez\ were calculated as the mean value of $VRI$ or $gri$ measurements.

\begin{table}
\begin{center}
\caption{New  TESS precise times of primary and secondary eclipses of \vv.}
\label{t1}
\smallskip
\begin{tabular}{lccccc}
\hline\hline\noalign{\smallskip}
BJD --     & Epoch & BJD --    &  Epoch    \\
24 00000   &       & 24 00000  &           \\ 
\hline\noalign{\smallskip}
 {\it Sector 19} &          & {\it Sector 40} &            \\  
  58816.30442  &  16385.0   &   59390.73118  &  17667.0 &   \\
  58816.52814  &  16385.5   &   59390.95830  &  17667.5 &   \\
  58817.64862  &  16388.0   &   59395.21180  &  17677.0 &   \\
  58818.99290  &  16391.0   &   59395.43938  &  17677.5 &   \\
  58826.83337  &  16408.5   &   59395.65992  &  17678.0 &   \\
  58830.86568  &  16417.5   &   59400.58914  &  17689.0 &   \\
  58831.09045  &  16418.0   &   59400.81648  &  17689.5 &   \\
  58831.31380  &  16418.5   &   59409.99947  &  17710.0 &   \\
  58831.53838  &  16419.0   &   59410.22612  &  17710.5 &   \\
  58840.72242  &  16439.5   &   59418.06483  &  17728.0 &   \\
  58840.94787  &  16440.0   &   59418.29167  &  17728.5 &   \\
 {\it Sector 20} &          &   59418.51314  &  17729.0 &   \\
  58842.74013  &  16444.0   &   59418.73930  &  17729.5 &   \\
  58842.96261  &  16444.5   &  {\it Sector 47}  \\                      
  58845.87647  &  16451.0   &   59580.04531  &  18089.5  &   \\
  58846.09935  &  16451.5   &   59580.26636  &  18090.0  &   \\ 
  58849.68331  &  16459.5   &   59585.64355  &  18102.0  &   \\ 
  58849.90907  &  16460.0   &   59588.55799  &  18108.5  &   \\ 
  58850.35711  &  16461.0   &   59590.12448  &  18112.0  &   \\ 
  58850.57951  &  16461.5   &   59592.14277  &  18116.5  &   \\ 
  58856.85193  &  16475.5   &   59592.81277  &  18118.0  &   \\ 
  58857.07832  &  16476.0   &   59595.72704  &  18124.5  &   \\ 
  58860.21499  &  16483.0   &   59595.94930  &  18125.0  &   \\ 
  58860.43710  &  16483.5   &   59598.63754  &  18131.0  &   \\ 
  58868.28071  &  16501.0   &   59602.89581  &  18140.5  &   \\
  58868.50233  &  16501.5   &   59603.11833  &  18141.0  &   \\ 
  58868.72892  &  16502.0   &   59606.48101  &  18148.5  &   \\ 
  --           &  --        &   59606.70294  &  18149.0  &   \\ 
  \hline\noalign{\smallskip}
\end{tabular}
\end{center}
\end{table}

\begin{table}
\begin{center}
\caption{New  and collected times of primary and secondary eclipses of \vv.}
\label{t1a}
\smallskip
\begin{tabular}{lccccc}
\hline\hline\noalign{\smallskip}
BJD --     & Epoch & Error   &  Observer/    \\
24 00000   &       & [day]   &  Source   \\
\hline\noalign{\smallskip}
  51474.62908 &      0.0  &  --      & Hoffmann et al. (2008) \\
  55478.61116 &   8936.0  &  0.0002  & Dubovsk\'y \\
  56356.38667 &  10895.0  &  0.0004  & H\"ubscher (2013) \\
  56371.62247 &  10929.0  &  0.0018  & H\"ubscher (2013) \\
  57080.47168 &  12511.0  &  0.0008  & H\"ubscher (2016) \\
  57296.43789 &  12993.0  &  0.0007  & H\"ubscher (2017) \\
  57815.30651 &  14151.0  &  0.0011  & Pagel (2018) \\
  58083.25282 &  14749.0  &  0.0001  & Screech  \\
  58115.29208 &  14820.5  &  0.0003  & Screech  \\
  58138.36421 &  14872.0  &  0.0003  & Screech  \\
  58174.43909 &  14952.5  &  0.0005  & Screech  \\
  58174.65936 &  14953.0  &  0.0001  & Screech  \\
  58245.45848 &  15111.0  &  0.0001  & Screech  \\
  58418.41320 &  15497.0  &  0.0001  & Screech  \\
  58506.45857 &  15693.5  &  0.0004  & Dienstbier \\
  58764.54982 &  16269.5  &  0.0004  & Pagel (2021)\\
  58887.32392 &  16543.5  &  0.0017  & Pagel (2021)\\
  58887.54772 &  16544.0  &  0.0013  & Pagel (2021)\\
  58957.44818 &  16700.0  &  0.0002  & 1 \\
  58961.48087 &  16709.0  &  0.0001  & 1 \\
  59107.55265 &  17035.0  &  0.0001  & Screech  \\
  59110.46119 &  17041.5  &  0.0002  & 1 \\
  59113.59834 &  17048.5  &  0.0002  & 1 \\
  59275.35892 &  17409.5  &  0.0094  & Pagel (2022)\\
  59275.58022 &  17410.0  &  0.0022  & Pagel (2022)\\
  59465.55996 &  17834.0  &  0.0002  & 2 \\
  59466.45638 &  17836.0  &  0.0001  & 2 \\
  59663.38181 &  18275.5  &  0.0005  & 2 \\
  59663.60753 &  18276.0  &  0.0001  & 2 \\
  59664.27982 &  18277.5  &  0.0003  & 2 \\
  59664.50386 &  18278.0  &  0.0001  & 2 \\
  59666.51932 &  18282.5  &  0.0003  & 2 \\
  59681.30668 &  18315.5  &  0.0008  & Kamenec \\
  59684.44221 &  18322.5  &  0.0008  & Kamenec \\ 
  59727.45761 &  18418.5  &  0.0001  & 1 \\
\hline
\end{tabular}

Observatory: 1 - \stef, 2 - \valmez
\end{center}
\end{table}

\section{O-C diagram}

\begin{figure}[t]
\centering
\includegraphics[width=0.75\textwidth]{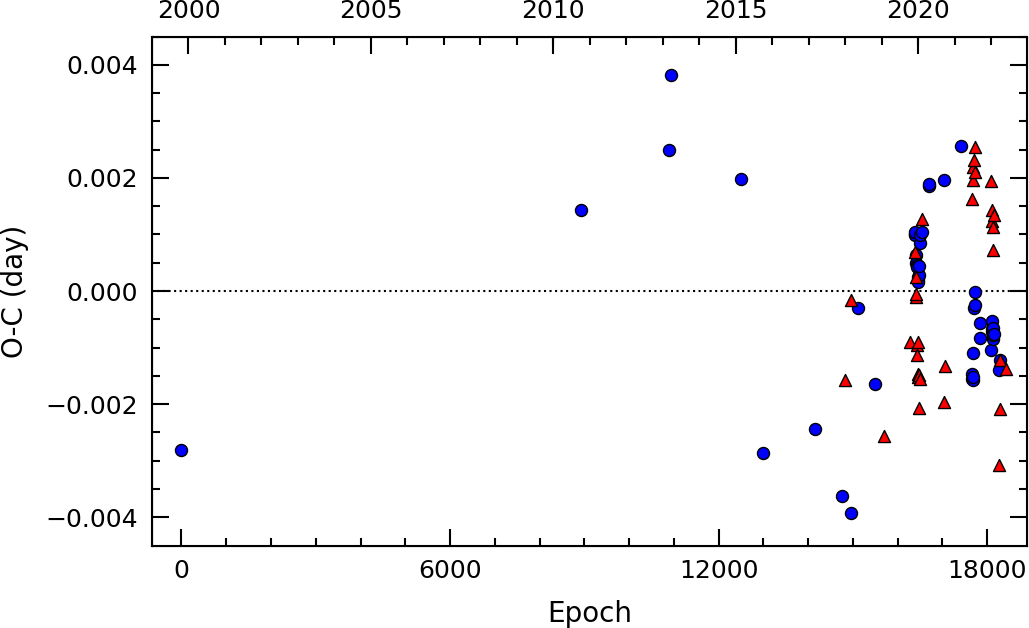}
\caption[ ]{The current \oc\ diagram for the times of minimum of \vv\ since discovery. The individual primary and secondary CCD minima are denoted by blue circles and red triangles, respectively.}
\label{608oc}
\medskip
\includegraphics[width=0.75\textwidth]{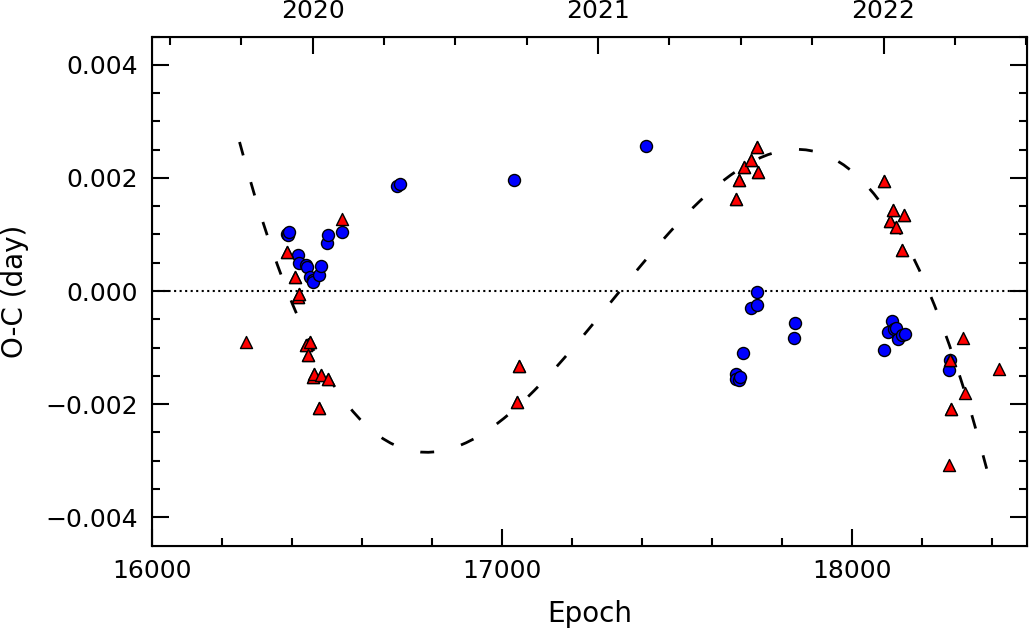}
\caption[ ]{The \oc\ diagram in detail for the times of minimum of \vv\ since the first {\sc Tess} data. The change of  \oc\ values for primary and secondary eclipses very probably caused by stellar surface activity is clearly visible. The quasi-sinusoidal fit for secondary minima with a semi-amplitude of about 0.003~day is plotted as the dashed black curve.  }
\label{608ocd}
\end{figure}

The period changes of \vv\ had not been studied since its discovery.
Only a few eclipse timings have been reported in the literature. 
Using the {\sc Tess} data we derived 54 new mid-eclipse times 
to complete our current \oc\ diagram. These times divided into four groups according to different {\sc Tess} sectors are included in Table~\ref{t1}.
Their uncertainty was fixed to 0.00001~day although the least squares fit of the light curve during the eclipse gives us formally smaller values. On the other hand, the scatter of {\sc Tess} minima is surprisingly larger than the given error. 
Besides those minima given in Table~\ref{t1}, we also used previous times 
of minimum published by H\"ubscher (2013, 2016, 2017) and  Pagel (2018, 2021, 2022),
and those collected in the \oc\ Gateway\footnote{\url{http://var2.astro.cz/ocgate/}}. 

Additional minima were obtained from photometric data available at the BAA Photometry Database\footnote{\url{https://britastro.org/photdb/data.php}}. 
These observations were carried out by James T. Screech using the 0.07-m refractor and CMOS Camera ASI 1600MM-C.  Five additional minima have been obtained by Pavol Dubovsk\'y using 400 mm photo-lens and CCD camera SXVF-H9, by Vojtěch Dienstbier using the 0.15-m Newtonian telescope and CCD camera MI G2-1600, by Mat\'u\v{s} Kamenec using the SkyWatcher 102/500 telescope and the Canon EOS 500D camera and finally again at Štefánik Observatory using 0,4-m Schmidt-Cassegrain telescope and the SBIG ST10-XM CCD camera.


Because the {\sc Tess} data are provided in the Barycentric Julian Date Dynamical Time (BJD$_{\rm TDB}$), all our times in Table~\ref{t1} were first transformed to this time scale using the often used Time Utilities of Ohio State University\footnote{\url{http://astroutils.astronomy.ohio-state.edu/time/} } 
(Eastman et al. 2010). 
A total of  88 precise CCD times including  37 secondary eclipses were used for the period determination. The computed linear light elements and their internal errors 
of the least-squares fit are given in Eq.~\ref{ephem},
the historical \oc\ diagram is shown in Fig.~\ref{608oc}, 
the \oc\ diagram for the current mid-eclipse times in detail is plotted 
in  Fig.~\ref{608ocd}. 

As one can see, the current accurate mid-eclipse times  show a relatively large scatter.
There are no clear indications for regular cyclical changes of the orbital period caused f.e. by a third component and observed as LITE.
Moreover, the shallower secondary eclipses show some cyclic trend  of their \oc\ values because they are probably more sensitive to some surface inhomogenities of the secondary component, see Fig.~\ref{608ocd}. 
This effect was further studied by a light-curve solution.
The observable difference in mid-eclipse times of spotted components 
was also outlined recently by Korda et al. (2017).

\section{Light curve solution}

\begin{figure}
\begin{center}
\includegraphics[width=0.95\textwidth]{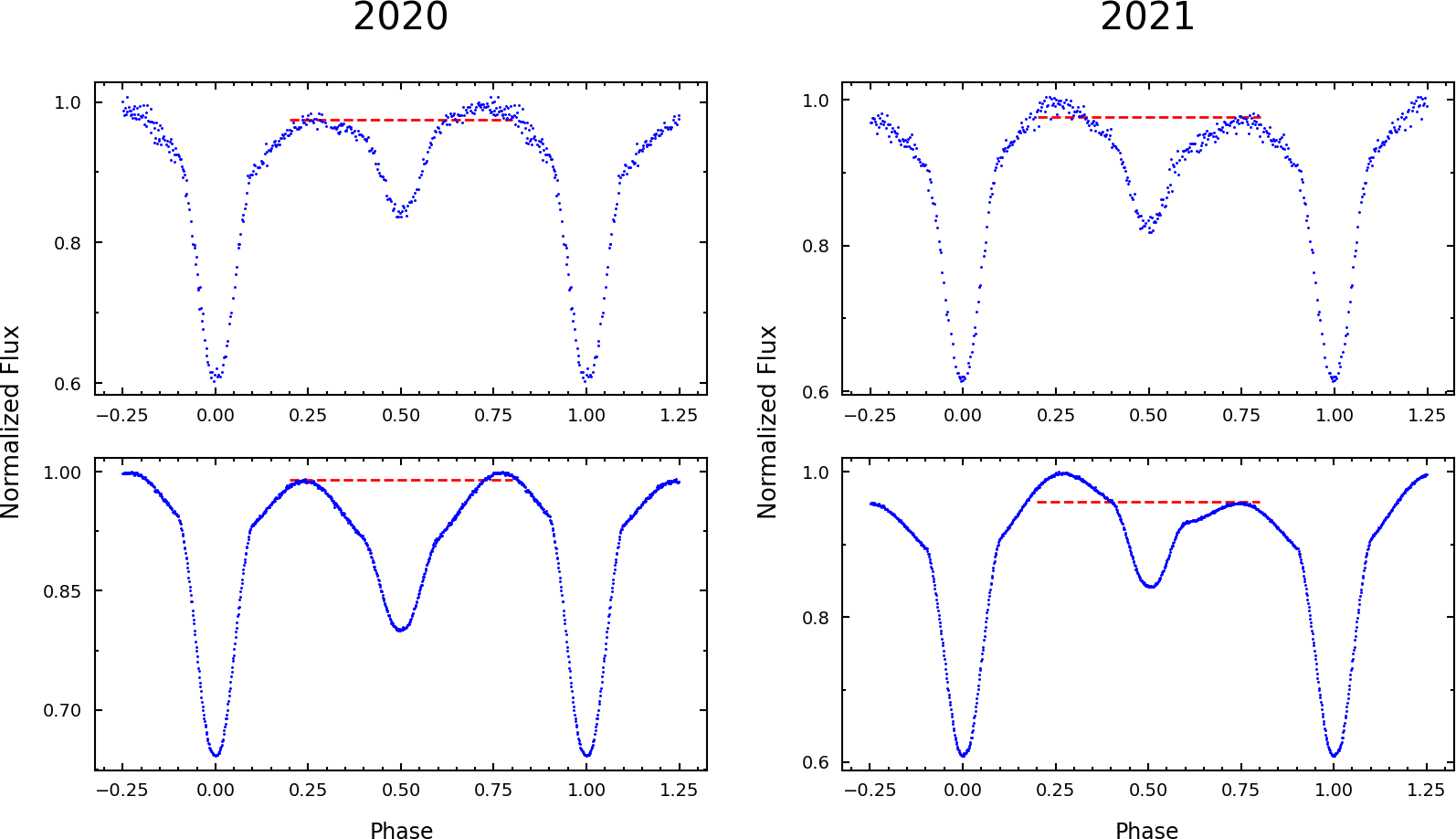}
\caption[ ]{Comparison of light curves obtained in 2020 and 2021. The first row corresponds to the light curves obtained at \stef\ and at \valmez, the second row corresponds to the {\sc Tess} light curves, sectors 20 and 40. Changes of brightness in both quadratures are denoted by red dashed lines.}
\label{compLC}
\end{center}
\end{figure}

\begin{figure}
\begin{center}
\includegraphics[width=0.6\textwidth]{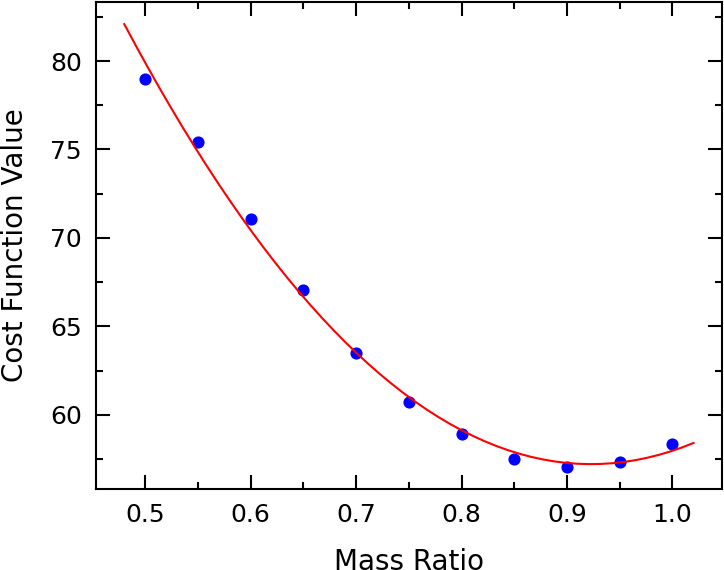}
\caption[ ]{The result of the $q$-search for the {\sc Tess} light curve. }
\label{608q}
\end{center}
\end{figure}

\begin{figure}
\begin{center}
\includegraphics[width=0.85\textwidth]{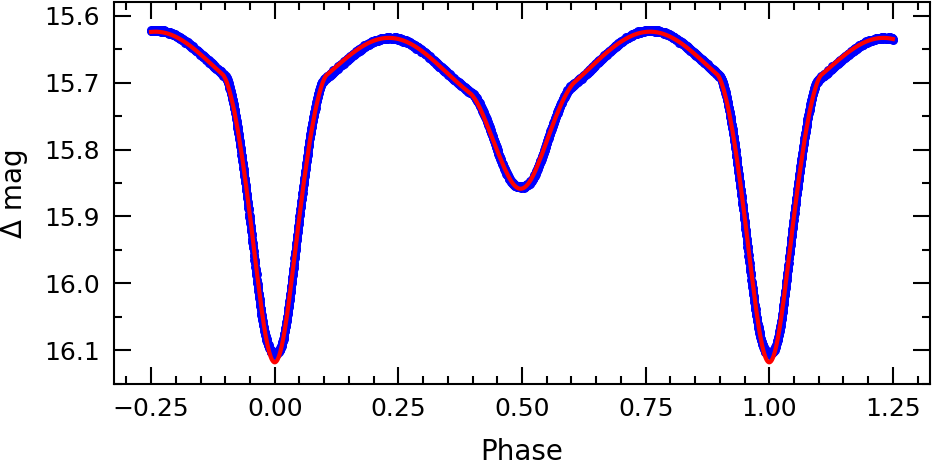}
\caption[ ]{The {\sc Phoebe} final solution for the {\sc Tess} light curve, Sector~20. The {\sc Tess} data are denoted blue, the resulting model in red. }
\label{608TESS}
\end{center}
\end{figure}

No photometric or spectroscopic solution has been reported for \vv\ so far. 
Comparison of our light curves obtained in different epochs 
with the {\sc Tess} data clearly show rapid changes of their shape, 
especially out of eclipse around phases 0.25 and 0.75, see Fig.~\ref{compLC}. 
All light curves of \vv\ obtained during 2020--2021 at our disposal were analyzed using the well-known {\sc Phoebe}\footnote{\url{http://phoebe-project.org/}} code (Pr\v{s}a \& Zwitter 2005, Pr\v{s}a et al. 2016), which is based on the Wilson-Devinney algorithm (Wilson \& Devinney 1971) and is widely used for modeling the photometric light curves of eclipsing binaries as a standard tool. Because \vv\ belongs to late-type binaries, we adopted the bolometric albedos and gravity darkening coefficients as $A_1 = A_2 = 0.5$ and $g_1 = g_2 = 0.32$, which corresponds to the convective envelopes. Synchronous rotation for both components of the system ($F_1 = F_2 = 1$) and a circular orbit ($e = 0$) were assumed. We used the logarithmic limb-darkening law with the coefficients adopted from van Hamme (1993) tables.
The temperature of the primary component was fixed to the value of 5~300~K given in the {\sc Gaia} DR2 Archive. This value is in good agreement with the color index $B-V = 0.872$ mag and $J-H = 0.403$ mag given in the {\sc Simbad} database. Also the table of Pecaut \& Mamajek (2013)\footnote{\url{http://www.pas.rochester.edu/~emamajek/EEM\_dwarf\_UBVIJHK\_colors\_Teff.txt}} gives for this temperature the similar color indexes $B-V = 0.816$ mag, $J-H = 0.387$ mag and the same spectral type K0.

In the absence of the spectroscopic mass ratio, the $q$-search process
was performed to find a corresponding photometric mass ratio. 
Assuming a given mass ratio $q$ from the interval of (0.5,1) in step of 0.05, we let converge the other parameters. At first we fitted those parameters that had a substantial effect on the shape of light curves.
The adjustable parameters were thus the inclination $i$, the effective temperature of the secondary component $T_2$, luminosity and dimensionless potentials of both components $\Omega_1, \Omega_2$. Next, the characteristics of a cold spot/region on the secondary component (colatitude, longitude, spot radius and temperature factor) were included. 
This method of the direct mass ratio and spot parameters estimation from the eclipsing binary light curve was recently evaluated by Terrell (2022).
Because the effective temperature and the radius of a spot are strongly correlated, we assumed that the ratio of spot/star temperatures is 
close to 0.9 in our analyzes.
No third light $L_3$ was included to the light curve solution. 
The fine and coarse grid raster for both components were set to 30.
The minimum value of the cost function was achieved at $q = 0.92 \pm 0.07$, see also Fig.~\ref{608q}. 


As a first attempt we selected roughly 1000 points of the precise {\sc Tess} light curve obtained in the Sector~20, to obtain basic photometric parameters of the system, see Fig.~\ref{608TESS} and Table~\ref{t3}. The wavelength coverage of {\sc Tess} is about 6000 -- 10000 \AA, covering most of the R and I filters. 
As one can see, the agreement between the theoretical and observed light curve is very good. 
Both additional light curves obtained in \stef\ and \valmez\ were then solved
independently to estimate primarily the characteristics of the dark region. 
All of available light curves in different filters obtained in one season were fitted simultaneously.
Numerous {\sc Phoebe~1} runs in a detached mode using the different setup of initial parameters were evaluated. The results as well as the cost function value were recorded.

The need to include spots to the final solution was evident in view of the modulation of out-of-eclipse light curves. We also made some preliminary tests placing spot on both components at latitudes of 45 deg, which is the region most affected by spots in low-mass stars (Granzer et al. 2000).

Finally, for a given temperature, the mass of the primary component 
$M_1 = 0.88$ \ms\ was adopted. Then the semi-major axis 
$a = 2.935 $ \rs\ was fixed to an appropriate value for the primary mass to be equal to a typical mass of a particular spectral type (Pecaut \& Mamajek 2013). With this approach, we were able to estimate the preliminary masses, in addition to the radii and luminosities of both components in absolute units (see Table~\ref{t5}).

\begin{figure}
\begin{center}
\includegraphics[width=0.95\textwidth]{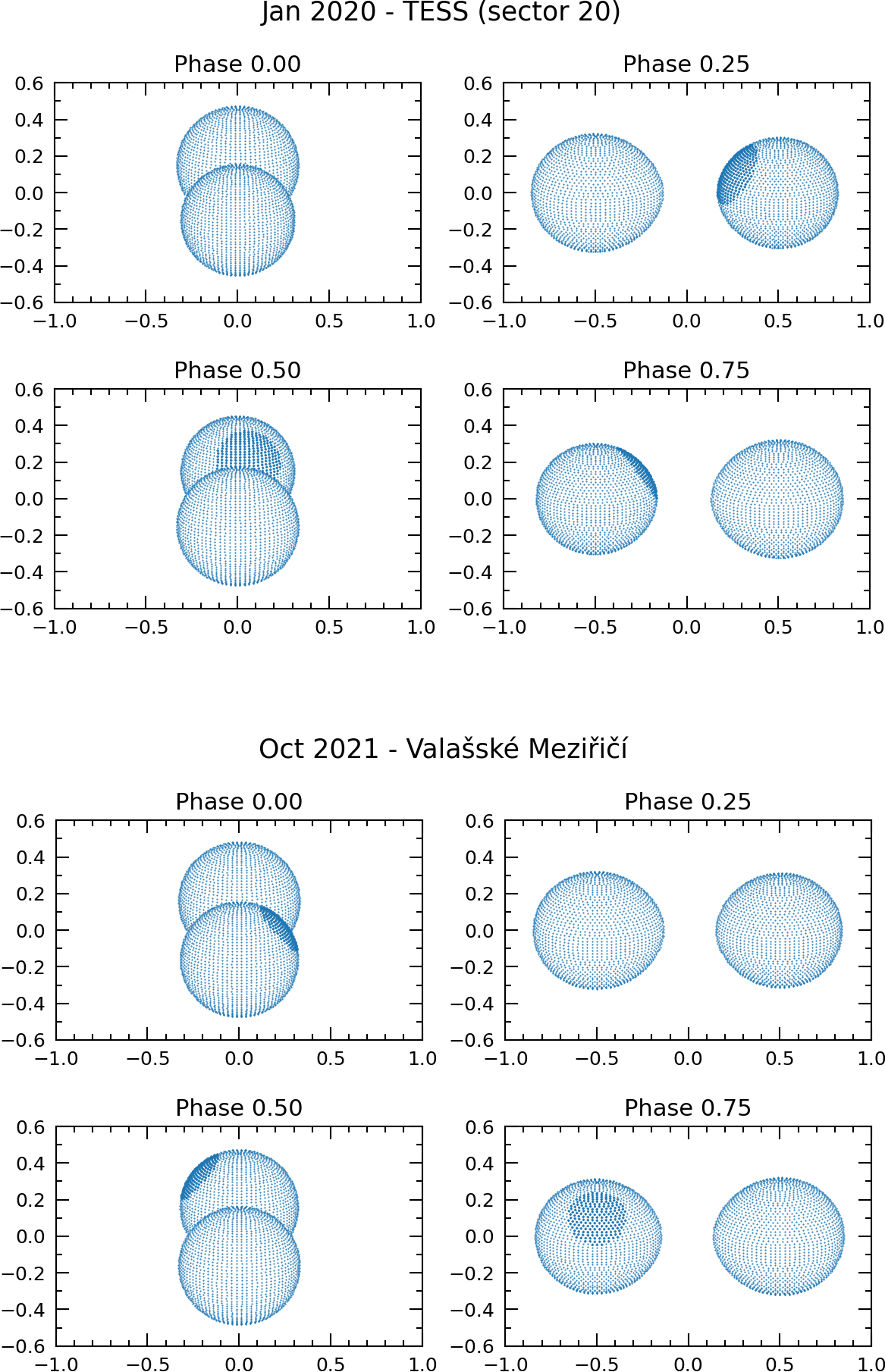}
\caption[ ]{The geometry of \vv\ in two epochs.  
The detached configuration with the large cold region on the secondary component is visible in the {\sc Tess} data as well as in our photometry obtained at \valmez.}
\label{608model}
\end{center}
\end{figure}

The final light curve solution is given in Table~\ref{t3},
where also bolometric limb-darkening coefficients and relative radii of both components are given.
The resulting parameters of the cold region on the secondary are given in Table~\ref{t7}. 
The geometrical representation of \vv\  in two epochs is displayed in Fig.~\ref{608model}.

\begin{table}
\begin{center}
\caption{The photometric elements of \vv, the {\sc Phoebe} solution.  }
\smallskip
\label{t3}
\begin{tabular}{ccc}
\hline\hline\noalign{\smallskip}
Parameter   & Primary & Secondary \\
\noalign{\smallskip}\hline\noalign{\smallskip}
$i$ [deg]       &  \multicolumn{2}{c}{73.8}\\
$q$             &  \multicolumn{2}{c}{0.92  (fixed)}\\
$T_{1,2}$  [K]  &  5300 (fixed) &  4110   \\
$\Omega_{1,2}$  &  3.879        &  4.256  \\
$X_{1,2}$       &  0.703        &  0.735    \\
$r(pole)$   &  0.333         &  0.284  \\
$r(side)$   &  0.346         &  0.291  \\
$r(point)$  &  0.387         &  0.309 \\
$r(back)$   &  0.365         &  0.302  \\
\noalign{\smallskip}\hline
\end{tabular}
\end{center}
\end{table}

\section{Discussion}

\begin{table}
\begin{center}
\caption{Absolute parameters of \vv, based on {\sc Tess} photometry
and color indexes.}
\smallskip
\label{t5}
\begin{tabular}{ccc}
\hline\hline\noalign{\smallskip}
Parameter         & Primary       & Secondary \\
\noalign{\smallskip}\hline\noalign{\smallskip}
Mass [\ms]         &    0.88    &   0.81(6)    \\
Radius [\rs]       &    1.02(4) &   0.86(7)    \\
Luminosity [L$_{\odot}$]  &  0.74(6)   &     0.19(3)      \\
$M_{\rm bol}$ [mag]&    5.06(9)    &   6.55(19)    \\
$a$ [\rs]          &  \multicolumn{2}{c}{2.94(3)} \\     
\noalign{\smallskip}\hline
\end{tabular}
\end{center}
\end{table}

In Table~\ref{t5} we collected absolute parameters of \vv\ directly derived from the light curves or estimated from the color indexes.
The mass of primary component was taken from the table of Pecaut \& Mamajek (2013) according to its {\sc Gaia} temperature. The other quantities in Table~\ref{t5} were derived directly from the light curve solution in {\sc Phoebe}. 

Moreover, for the {\sc Gaia} parallax given in Table~\ref{tg} and the angular diameter of the system $\rho = 7.84 \cdot 10^{-2}$ mas (Cruzalebes et al. 2019), one can obtain for the distance of \vv\ a linear dimension $D$ = 1.96 \rs, which is in good accordance with the derived values of radii $R_1 + R_2 = 1.88$ \rs\ (see Table~\ref{t5}). No flare-like event was recorded during our photometric monitoring as well as on the precise {\sc Tess} light curves.

\begin{figure}
\begin{center}
\includegraphics[width=0.85\textwidth]{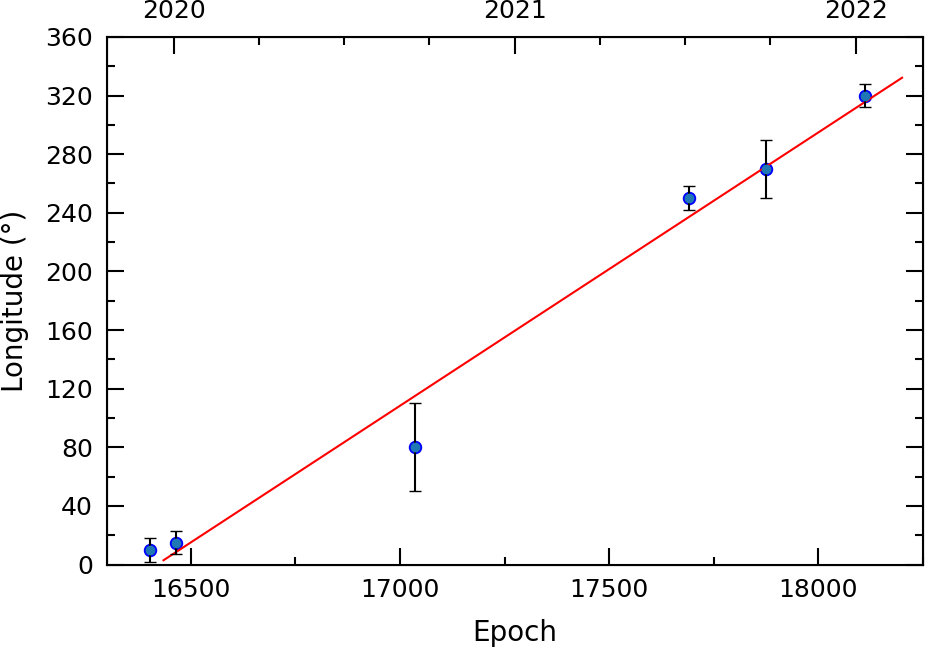}
\caption[ ]{The gradual increase in longitude of the cold structure on the surface of the secondary component during several observational epochs. The calculated slope of the red line (0.185 deg/epoch) corresponds to a rate of 12.5 deg/month. Mean errors in longitude are indicated.}
\label{608spot}
\end{center}
\end{figure}

On the other hand, the seasonal light curve of \vv\ shows slowly evolving starspot structures that also strongly affect the estimation of precise \oc\ timings.
The apparent period changes are better visible on the run of secondary eclipses and are probably caused by a moving cold structure on the surface of the secondary component, where a different part of this region is partially eclipsed.
 Fig.~\ref{608spot} shows a slow increase in the spot longitude during several epochs of {\sc Tess} and our observations collected 
also in Table~\ref{t7}. Although the resulting colatitude and the radius of the dark structure remain practically the same, the value 
of its longitude changes at a mean rate about +12.5 deg/month. This value is in good agreement with the current \oc\ diagram (Fig.~\ref{608ocd}), where secondary minima show a quasi-periodic behavior with a period about 2.4 years.

\begin{table}
\begin{center}
\caption{ Parameters of the cold surface structure on the secondary component of \vv\ obtained using different data sets.}
\smallskip
\label{t7}
\begin{tabular}{lcccccc}
\hline\hline\noalign{\smallskip}
Data set   & Epoch & Longitude & Colatitude & Radius & {\sc Phoebe} \\
           &       & [deg]     &  [deg]     & [deg] & cost function \\
\noalign{\smallskip}\hline\noalign{\smallskip}
TESS 19	       & Dec 2019  &  10 & 70  & 35 & 108\\
TESS 20	       & Jan 2020  &  15 & 60  & 35 &  43 \\
\v{S}tef. Obs. & Sep 2020  &  80 & 70  & 30 & 381\\
TESS 40	       & Jul 2021  & 250 & 70  & 35 & 118 \\
Val.Mez.Obs.   & Oct 2021  & 270 & 50  & 30 & 302 \\
TESS 47        & Jan 2022  & 320 & 70  & 35 &  52 \\
\noalign{\smallskip}\hline
\end{tabular}
\end{center}
\end{table}


The similar result was obtained recently  for the binary V1130~Cyg, where the active component has one spotted area migrating longitudinally on the stellar surface with a period of about 122~days (Yoldas \& Dal, 2021).
Star-spot activity was also studied in case of KIC~11560447 by Ozavci et al. (2018), where the rotation of active regions up to about 2.4 deg per day was found.
 The spot activity in Algol binary KIC~06852488 was recently presented by Shi et al. (2021).
Using the Kepler and {\sc Tess} light curves they found an evolving hot spot on the primary and a cold spot on the secondary component. Certain correlation between the O’Connell effect and the \oc\ curve similar to our result was presented.  

One has to take into consideration that such small apparent variation in orbital period on a short time-scale, frequently attributed to the light-time effect caused by an unseen third body, could be simply a result of the dark-spot evolution or its movement on the surface of binary components.


\section{Conclusions}

A study of late-type and low-mass binaries provides us with important information about the most common stars in our Galaxy. The interesting system \vv\ with a large cold region on the surface of the secondary component was studied. 
The new multi-color photometric observations of \vv\ as well as {\sc Tess} data were used to investigate its light curve and determine the photometric parameters of its components.

The system \vv\ contains the more massive and warmer primary component 
with $M_{\rm 1} = 0.88$ \ms.  The mass of the secondary was determined 
$M_{\rm 2} = 0.81 \pm 0.06$~\ms\ and its
effective temperature $T_2 = 4110 \pm 50$~K. The system inclination is $i=73.8^\circ$.
We can conclude that \vv\ is a nearby low-mass, active and detached eclipsing binary.

Their apparent period changes are probably connected with magnetic activity and 
changes of position of the cold structure on the secondary component.
The current \oc\ diagram based primarily on {\sc Tess} data shows quasi-periodic changes caused by surface activities and a systematic shift of secondary eclipses. Formally one could declare such behavior as a "false eccentricity" of the otherwise circular orbit. This phenomenon must also be included to possible explanations of period changes in many late-type binary systems explained so far by a magnetic field (so-called Applegate mechanism, Applegate 1992) or by the light-time effect (LITE) caused by a third component orbiting the eclipsing pair (Irwin 1952).
Concerning the surface activity, the identification of \vv\ as a faint X-ray source (Voges et al. 1999) also supports our findings. 

New high-accuracy timings of this eclipsing binary are necessary
to investigate the parameters derived in this paper. 
It is also highly desirable to obtain new, high-dispersion
and high-S/N spectroscopic observations and to apply modern disentangling methods to obtain the radial-velocity curves and to derive accurate masses for this interesting late-type system. 

\section{Acknowledgments}
\small{
The authors would like to thank Ladislav \v{S}melcer, \valmez\ observatory, and Zbyn\v{e}k Henzl, Variable Star and Exoplanet Section, the Czech Astronomical Society, for their kind assistance with remote observation, and to Stanislav Boula and other employees of \stef\ for their support. 
The research of M.W. was supported by the project Progress Q47 {\sc Physics} of Charles University in Prague.
This paper includes data collected by the {\sc Tess} mission. 
Funding for the {\sc Tess} mission is provided by the NASA's Science Mission Directorate.
This work has made use of data from the European Space Agency (ESA) mission 
{\sc Gaia}\footnote{\url{https://www.cosmos.esa.int/gaia}}, processed 
by the {\sc Gaia} Data Processing and Analysis Consortium (DPAC) 
\footnote{\url{https://www.cosmos.esa.int/web/gaia/dpac/consortium}}. 
Funding for the DPAC has been provided by national
institutions, in particular the institutions participating in the Gaia Multilateral Agreement. 
The following internet-based resources were used in research
for this paper: the SIMBAD and VIZIER database operated at CDS, Strasbourg, France, the NASA's Astrophysics Data System Bibliographic Services, the International Variable Star Index (VSX) database, operated at AAVSO, Cambridge, Massachusetts, USA,  the British Astronomical Association, Photometry Database, and the Czech Astronomical Society B.R.N.O. photometry database of eclipsing binary stars minima and associated O-C Gateway database of minima timings.
This research is part of an ongoing collaboration between professional astronomers, the Czech Astronomical Society, Variable Star and Exoplanet Section and
the Planetum Prague, \stef.
}

\section{Data Availability}

The {\sc Tess} and {\sc Gaia} data underlying this paper is publicly available,
the other photometric data are available on reasonable request to the authors.

\end{document}